\documentclass{ws-procs9x6-cpt25}
\usepackage{slashed}
\usepackage{upgreek}
\usepackage{units}
\usepackage{booktabs}
\begin{document}
	
	\newcommand{\refeq}[1]{(\ref{#1})}
	\def\etal {{\it et al.}}
	
	\title{Vacuum Cherenkov Radiation for\\
		Nonminimal Isotropic Lorentz Violation}
	
	\author{Albert Yu.\ Petrov,$^1$ Marco Schreck,$^{2,3}$ and Alexandre R.\ Vieira$^4$}
	
	\address{$^1$Departamento de F\'isica, Universidade Federal da Para\'iba, Caixa Postal 5008,\\
		Jo\~ao Pessoa 58051-970, Para\'iba, Brazil}
    \address{$^2$Programa de P\'{o}s-gradua\c{c}\~{a}o em F\'{i}sica, Universidade Federal do Maranh\~{a}o,\\
        Campus Universit\'{a}rio do Bacanga, S\~ao Lu\'is 65080-805, Maranh\~{a}o, Brazil}
    \address{$^3$Coordena\c{c}\~{a}o do Curso de F\'{i}sica -- Bacharelado, Universidade Federal do Maranh\~ao,\\
        Campus Universit\'{a}rio do Bacanga, S\~ao Lu\'is 65080-805, Maranh\~{a}o, Brazil}
	\address{$^4$Departamento de Ciências Exatas e Educaç\~ao, Instituto de Ci\^encias Agr\'arias, Exatas e Biol\'ogicas de Iturama -- ICAEBI,\\
		Universidade Federal do Tri\^angulo Mineiro -- UFTM,\\
		Iturama, Minas Gerais 38280-000, Brazil}
	
	\begin{abstract}
		In this work, we study the effects of vacuum Cherenkov radiation caused by nonminimal dimension-5 Lorentz-violating (LV) operators in the fermion sector. Explicitly, we focus on two independent isotropic pieces of each set of nonminimal coefficients. Under the assumption that vacuum Cherenkov radiation is an expected phenomenon, experimental data of ultra-high-energy cosmic rays (UHECRs) allow us to put stringent bounds on isotropic coefficients in quarks.
	\end{abstract}
	
	\bodymatter
	
	\section{Introduction}
	\label{Sec1}
	
	Lorentz and CPT symmetries play a fundamental role in local relativistic field theories. If these symmetries are broken at the Planck scale, a multitude of effects can arise in the low-energy effective description via field theories. This concerns particle phenomenology, on the one hand, and gravitational waves, on the other hand. Prominent phenomena are shifts of energy levels of electrons in bound systems, anisotropies in particle and wave propagation, vacuum birefringence, and unusual processes such as Cherenkov-type radiation \textit{in vacuo}. As a consequence, precise measurements of particle and gravitational-wave propagation can constrain several sets of Lorentz- and CPT-violating coefficients, testing to which extent these symmetries are exact in nature.
	
	Cherenkov radiation occurs when a massive, charged particle travels through a medium with a velocity that exceeds
	the speed of light in this medium. After all, the latter is reduced by the refractive index of the material.
	At the same time, Lorentz symmetry breaking implies a nontrivial refractive index of the vacuum such that
	Cherenkov-type radiation \textit{in vacuo} can be expected under certain circumstances.
	Indeed, various studies of vacuum Cherenkov radiation within the Standard-Model Extension (SME) have been carried out to date.
	Most of these analyses are based on modifications of the photon sector in the minimal SME, such as the Carroll--Field--Jackiw term \cite{CSCr,Colladay:2016rmy} or CPT-even modified Maxwell theory.\cite{CPTevenCr} Vacuum Cherenkov radiation has also been investigated in the gravity sector,\cite{GravCr} for pions,\cite{pions} partons,\cite{partons} and quarks.\cite{Marco} These works have in common that they delve into the minimal SME, whereas a comprehensive analysis of nonminimal operators has not been carried out so far. In contrast to minimal coefficients, nonminimal ones are expected to become dominant at high energies. We intend to fill this gap partially by reporting on recent results on isotropic dimension-5 coefficients. Computational details will be given elsewhere.\cite{MainPaper}
	
	\section{Basics}
	\label{Sec2}
	
	We explore modified Dirac fermions coupled to Maxwell electrodynamics, where the modification is governed by nonminimal LV operators of the SME
	fermion sector:
	\begin{subequations}
		\label{eq:modified-QED}
		\begin{align}
			\mathcal{L}_{\mathrm{QED}}&=\mathcal{L}_{\upgamma}+\mathcal{L}_{\psi}\,, \\[1ex]
			\mathcal{L}_{\upgamma}&=-\frac{1}{4}F^{\mu\nu}F_{\mu\nu}-\frac{1}{2}(\partial_{\mu}A^{\mu})^2\,, \\[1ex]
			\mathcal{L}_{\psi}&=\frac{1}{2}\bar{\psi}(\gamma^{\mu}i\mathcal{D}_{\mu}-m_{\psi}+\widehat{\mathcal{Q}})\psi+\text{H.c.}\,,
		\end{align}
	\end{subequations}
	with the \textit{U}(1) gauge field $A_{\mu}$, the electromagnetic field strength tensor $F_{\mu\nu}=\partial_{\mu}A_{\nu}-\partial_{\nu}A_{\mu}$, the Dirac spinor field $\psi$, the Dirac conjugate field $\bar{\psi}:=\psi^{\dagger}\gamma^0$, and the fermion mass $m_{\psi}$. All fields are defined in Minkowski spacetime with the metric tensor $\eta_{\mu\nu}$ of signature $(+,-,-,-)$. The Dirac matrices $\gamma^{\mu}$ satisfy the Clifford algebra $\{\gamma^{\mu},\gamma^{\nu}\}=2\eta^{\mu\nu}$. The photon sector is minimally coupled to the fermion sector via the gauge-covariant derivative $\mathcal{D}_{\mu}$. The operator $\widehat{\mathcal{Q}}$ contains all nonminimal LV coefficients to be considered as follows.
	
	In the quantum regime, vacuum Cherenkov radiation is allowed when $\Delta E=0$, where $\Delta E$ is the energy balance equation given as follows:
	\begin{equation}
		\Delta E=E({\bf q})-|{\bf k}|-E({\bf q}-{\bf k})\,.
		\label{eqDR}
	\end{equation}
	Here, $E(\mathbf{q})$ denotes the modified fermion dispersion relation, where ${\bf q}$ and ${\bf k}$ are the momenta of the initial fermion and the standard photon emitted, respectively. The range of the final-photon momentum is restricted by the solutions for the polar angle. Solving the energy balance equation leads to $\theta=\theta_0=\theta_0(q,k,X_{\subset})$, where $-1\leq \cos \theta_0\leq 1$, $q\equiv|\mathbf{q}|$, $k\equiv|\mathbf{k}|$, and $X_{\subset}$ is a subset of LV coefficients.
	
\begin{figure}[t]
		\centering
        \includegraphics[scale=0.75]{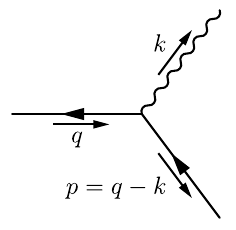}
		\caption{Tree-level Feynman diagram for vacuum Cherenkov radiation. There is an incoming fermion with momentum $\mathbf{q}$ and the outgoing photon and fermion with momenta $\mathbf{k}$ and $\mathbf{p}=\mathbf{q}-\mathbf{k}$, respectively.}
		\label{fig0}
	\end{figure}

	The decay rate $\Gamma$ is calculated at tree level, where the corresponding Feynman diagram is presented in Fig.~\ref{fig0}.
	The amplitude $\mathcal{M}$ reads
	\begin{equation}
		\mathcal{M}=e\bar{u}^{(s)}(q-k)\Gamma^{\mu} u^{(s')}(q)\epsilon_{\mu}^{(\lambda)}(k)\,,
	\end{equation}
	with the elementary charge $e$ as well as the indices $s$ and $\lambda$ for spin and polarization states, respectively. Moreover, $\Gamma^{\mu}$ is
	a generalized Dirac matrix that includes information on nonminimal SME coefficients. For example, $\Gamma^{\alpha}=\gamma^{\alpha}-(a^{(5)})^{\mu\beta\alpha}\gamma_{\mu}p_{\beta}$ for the CPT-odd operator $\hat{a}^{\mu}$. Gauge invariance can be checked by computing $k_{\mu}\mathcal{M}^{\mu}=0$ for $\mathcal{M}^{\mu}=e\bar{u}(q-k)\Gamma^{\mu} u(q)$.
	
	It is possible to show\cite{Marco} that this decay rate for isotropic frameworks can be written as:
	\begin{subequations}
		\begin{align}
			\Gamma&=\frac{1}{2E({\bf q})}\gamma\,, \\[1ex]
			\gamma&=\frac{1}{8\pi}\int_0^{k_{\mathrm{max}}}\mathrm{d}k\, \Pi(k)|\mathcal{M}|^2|_{\theta=\theta_0}\,, \\[1ex]
			\Pi(k)&=\frac{k\sin \theta}{E({\bf q}-{\bf k})}\left|\frac{\partial \Delta E}{\partial \theta}\right|^{-1}\Big|_{\theta=\theta_0}\,,
			\label{eq0.2}
		\end{align}
	\end{subequations}
	where $\Pi(k)$ is the phase space factor, and $|\mathcal{M}|^2$ is the amplitude squared averaged over the initial-fermion spin states and summed over the final-particle spin and polarization states. Moreover, $k_{\mathrm{max}}$ is the maximum possible value for the photon momentum, which is restricted by the requirement that $-1\leq \cos \theta_0\leq 1$.
	
	\section{Isotropic sector of dimension-5 operator $\hat{m}$}
	\label{Sec3}
	
	The nonminimal dimension-5 coefficients $m^{(5)}_{\alpha\beta}$ are CPT-even and possess two independent isotropic pieces, since the background tensor is not necessarily traceless. They are contained in the term
	\begin{equation}
		-\frac{1}{2}(m^{(5)})^{\alpha\beta}\bar{\psi}\mathrm{i}D_{(\alpha}\mathrm{i}D_{\beta)}\psi+\text{H.c.} \subset \mathcal{L}^{(5)}_{\psi D}\,,
	\end{equation}
	of Tab.~I in Ref.~\refcite{Zonghao}. The nonminimal LV operator within the modified Dirac equation in momentum space is given by\cite{Mewes}
	\begin{equation}
		\widehat{\mathcal{Q}}^{(5)}=-m^{(5)}_{\alpha\beta}p^{\alpha}p^{\beta}\supset -\mathring{m}_{0}^{(5)}(p^0)^2-\mathring{m}_{2}^{(5)}p^2\,,
		\label{eq1.0}
	\end{equation}
	where $\mathbf{p}^2\equiv p^2$, $\mathring{m}_0^{(5)}\equiv m^{(5)}_{00}$, and $\mathring{m}_{2}^{(5)}\equiv
	m^{(5)}_{11}=m^{(5)}_{22}=m^{(5)}_{33}$ are the isotropic parts. We will drop the superscript `$(5)$', for brevity.
	
	To find the dispersion relations, we look for nontrivial solutions of the modified Dirac equation $(\slashed{p}-m_{\psi}+\widehat{\mathcal{Q}})\psi=0$ and solve the determinant equation for $p_0$. First, there are four spurious dispersion relations:
	\begin{align}
		E&=\pm\frac{1}{ \sqrt{2}}\sqrt{\frac{1}{\mathring{m}_0^2}-\frac{2m_{\psi}}{\mathring{m}_0}+\frac{\sqrt{1-4m_{\psi}\mathring{m}_0-4\mathring{m}_0^2p^2}}{\mathring{m}_0^2}}
		\nonumber \\
		&=\pm\frac{1}{\mathring{m}_0}\mp m_{\psi} \mp (p^2+2m_{\psi}^2)\frac{\mathring{m}_0}{2}+\dots\,.
		\label{eq1.1}
	\end{align}
	In the Lorentz-invariant limit, these expressions do not reproduce the conventional dispersion relations for massive particles. Instead, they are singular. In principle, such spurious dispersion relations can be interpreted as describing Planck-scale effects, whereupon their meaning in effective field theory can be questioned. For this reason, and since our analysis is restricted to tree-level effects, Eq.~\eqref{eq1.1} will be ignored.
	
	The remaining four dispersion relations are perturbatively consistent and spin-degenerate. The positive-energy dispersion relation, which holds for particles, reads
	\begin{align}
		E_{\mathring{m}_0}&=\frac{1}{\sqrt{2}}\sqrt{\frac{1}{\mathring{m}_0^2}-\frac{2m_{\psi}}{\mathring{m}_0}-\frac{\sqrt{1-4m_{\psi}\mathring{m}_0-4\mathring{m}_0^2p^2}}{\mathring{m}_0^2}}
		\nonumber\\
		&=(m_{\psi}\mathring{m}_0+1) \sqrt{p^2+m_{\psi}^2}+\dots\,.
		\label{eq1.2}
	\end{align}
	Equation~(\ref{eq1.2}) enables us to solve the energy balance equation~\eqref{eqDR} for the polar angle
	$\theta$. Restricting ourselves to first order in the isotropic coefficient and employing $|\mathbf{q}|\equiv q$, we find
	\begin{equation}
		\cos \theta=\frac{1}{q}\left[\sqrt{q^2+m_{\psi}^2}+m_{\psi}\mathring{m}_0\left(k-\sqrt{q^2+m_{\psi}^2}\,\right)\right]\,.
		\label{eq1.3}
	\end{equation}
	The magnitude of the photon momentum has a maximum corresponding to the maximum value of
	$\cos \theta$. This maximum momentum to first order in $\mathring{m}_0$ is given by:
	\begin{align}
		k_{\mathrm{max}}&=\frac{q-\sqrt{q^2+m_{\psi}^2}}{ m_{\psi} \mathring{m}_0}+\frac{1}{4}m_{\psi} \mathring{m}_0 \left(\sqrt{q^2+m_{\psi}^2}+q\right)\nonumber\\
		&\phantom{{}={}}+\frac{1}{2} \left(3 q-\sqrt{q^2+m_{\psi}^2}\,\right)\,.
		\label{eq1.4}
	\end{align}
	Vacuum Cherenkov radiation takes place for positive $k$, where a minimum nonzero initial-fermion energy follows from the relation $k_{\mathrm{max}}=0$. Thus, the process occurs when the initial-fermion energy lies above a certain threshold given by
	\begin{equation}
		q_{\mathrm{th}}=\sqrt{\frac{m_{\psi}}{3\mathring{m}_0}}+\dots\,,
		\label{eq1.5}
	\end{equation}
	where we see that only positive values of the isotropic coefficient $\mathring{m}_0$ are permitted. It is observed that $q_{\mathrm{th}} \rightarrow \infty $ for $\mathring{m}_0 \rightarrow 0$, as expected, since the process is forbidden in the Lorentz-invariant limit. The phase space factor follows from the kinematics according to Eq.~\eqref{eq0.2}. In addition to the kinematic ingredients, we also need to take into account the dynamics of the process, which is described by the amplitude squared. Details on how to compute the latter will be provided in Ref.~\refcite{MainPaper}.
	
	The same procedure can be repeated for the other independent isotropic piece governed by $\mathring{m}_2$. Since the latter is not accompanied by time derivatives, spurious dispersion relations do not emerge. The perturbative positive-energy dispersion relation reads
	\begin{equation}
		E_{\mathring{m}_2}=\sqrt{m_{\psi}^2 +(1+ 2 m_{\psi} \mathring{m}_2) p^2 +  \mathring{m}_2^2 p^4}\,.
		\label{eq1.6}
	\end{equation}
	In this case, the polar angle and the maximum energy of the photon in terms of $q$ are complicated
	functions even at leading order in $\mathring{m}_2$. Nevertheless, the procedure for finding the threshold energy is the same as before. Doing so provides a result similar to Eq.~\eqref{eq1.5}:
	\begin{equation}
		q_{\mathrm{th}}=\sqrt{\frac{m_{\psi}}{3\mathring{m}_2}}+\dots\,.
	\end{equation}
	The coefficient $\mathring{m}_2$ must also be positive for the process to occur and the threshold is singular in $\mathring{m}_2$, as expected.
	\begin{figure}[t]
		\centering
		\includegraphics[scale=0.35]{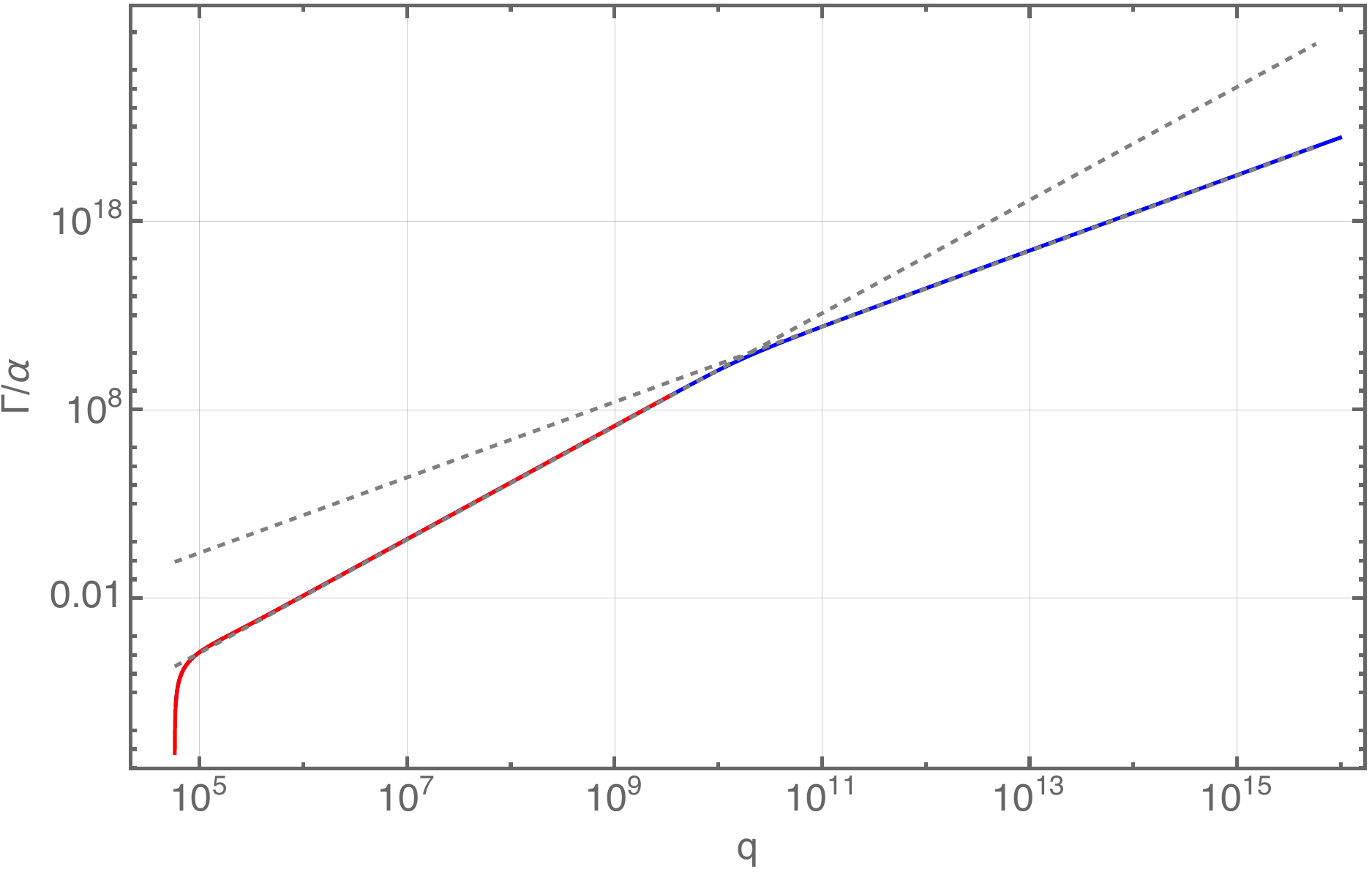}
		\caption{Double-logarithmic plot of the decay rates $\Gamma/\alpha$ with the fine-structure constant $\alpha=e^2/(4\pi)$ as functions of the initial-fermion momentum $q$ for $m_{\psi}=\unit[1.00]{GeV}$ as well as the isotropic coefficients $\mathring{m}_0=\unit[10^{-10}]{GeV^{-1}}$ and $\mathring{m}_2=\unit[10^{-10}]{GeV^{-1}}$, respectively. Both curves lie on top of each other.}
		\label{fig2}
	\end{figure}
	
	The decay rates for each of the two settings with either nonzero $\mathring{m}_0$ or $\mathring{m}_2$ are presented in Fig.~\ref{fig2}. In fact, both curves cannot be distinguished from each other. Several comments are in order. First, the equal threshold momenta can clearly be perceived, since $\Gamma\rightarrow 0$ when $q\rightarrow q_{\mathrm{th}}$, as expected. Second, the curve for $\mathring{m}_2$ possesses a feature resembling a knee, where the behavior of the decay rate changes drastically. Interestingly, a comparable behavior was observed in Ref.~\refcite{Colladay:2016rmy} for the vacuum Cherenkov process with modified photons. Third, the curve for $\mathring{m}_0$ stops before this knee, since Eq.~\eqref{eq1.2} becomes complex for momenta exceeding a certain upper limit.
	
	\section{Isotropic sector of dimension-5 operator $\hat{a}^{\mu}$}
	\label{Sec4}
	
	The nonminimal dimension-5 coefficients $a^{(5)}_{\alpha\beta\mu}$ are CPT-odd and have two independent isotropic pieces as does $m^{(5)}_{\mu\nu}$ studied previously. They are contained in the term
	\begin{equation}
		-\frac{1}{2}(a^{(5)})^{\mu\alpha\beta}\bar{\psi}\gamma_{\mu}\mathrm{i}D_{(\alpha}\mathrm{i}D_{\beta)}\psi+\text{H.c.}\subset \mathcal{L}^{(5)}_{\psi D}\,,
	\end{equation}
	of Tab.~I in Ref.~\refcite{Zonghao}. The operator in the modified Dirac equation reads $\widehat{\mathcal{Q}}=-(a^{(5)})^{\mu\alpha\beta}p_{\alpha}p_{\beta}\gamma_{\mu}$ such that the isotropic parts are formulated as follows:\cite{Mewes}
	\begin{equation}
		(a^{(5)})^{\mu\alpha\beta}p_{\alpha}p_{\beta}\gamma_{\mu} \supset \mathring{a}^{(5)}_0 p^{0}p^{0}\gamma^{0}+\frac{1}{3}\mathring{a}^{(5)}_2(p^{k}p^{k}\gamma^{0}+2p^0p^k\gamma^k)\,,
		\label{eq2.1b}
	\end{equation}
	where the superscript `$(5)$' will be omitted, for brevity.
	The dispersion relations follow again from solving the modified Dirac equation in momentum space. For the isotropic
	part $\mathring{a}_0$, the condition of a vanishing determinant of the Dirac operator implies four spurious dispersion relations, such as for~$\mathring{m}_0$. The perturbative dispersion relations are again spin-degenerate. The positive-energy result at first order in the isotropic
	coefficient is
	\begin{equation}
		E_{\mathring{a}_0}=\sqrt{p^2+m_{\psi}^2}+ \mathring{a}_0 \left(p^2+m_{\psi}^2\right)+\dots\,.
		\label{eq2.3}
	\end{equation}
	By solving the energy balance equation~\eqref{eqDR} with Eq.~\eqref{eq2.3}, we find the polar angle. As previously, the maximum photon momentum $k_{\mathrm{max}}$ follows from the maximum of $\cos \theta$. The threshold momentum of the initial fermion is computed from
	$k_{\mathrm{max}}=0$. Performing a computation exact in $\mathring{a}_0$ leads to
	\begin{equation}
		\label{eq:threshold-a0}
		q_{\mathrm{th}}=\sqrt[3]{\frac{m^2_{\psi}}{4\mathring{a}_0}}+\dots\,,
	\end{equation}
	where $q_{\mathrm{th}}\rightarrow\infty$ holds for $\mathring{a}_0\rightarrow 0$, as expected.
	\begin{figure}[t]
		\centering
		\includegraphics[scale=0.35]{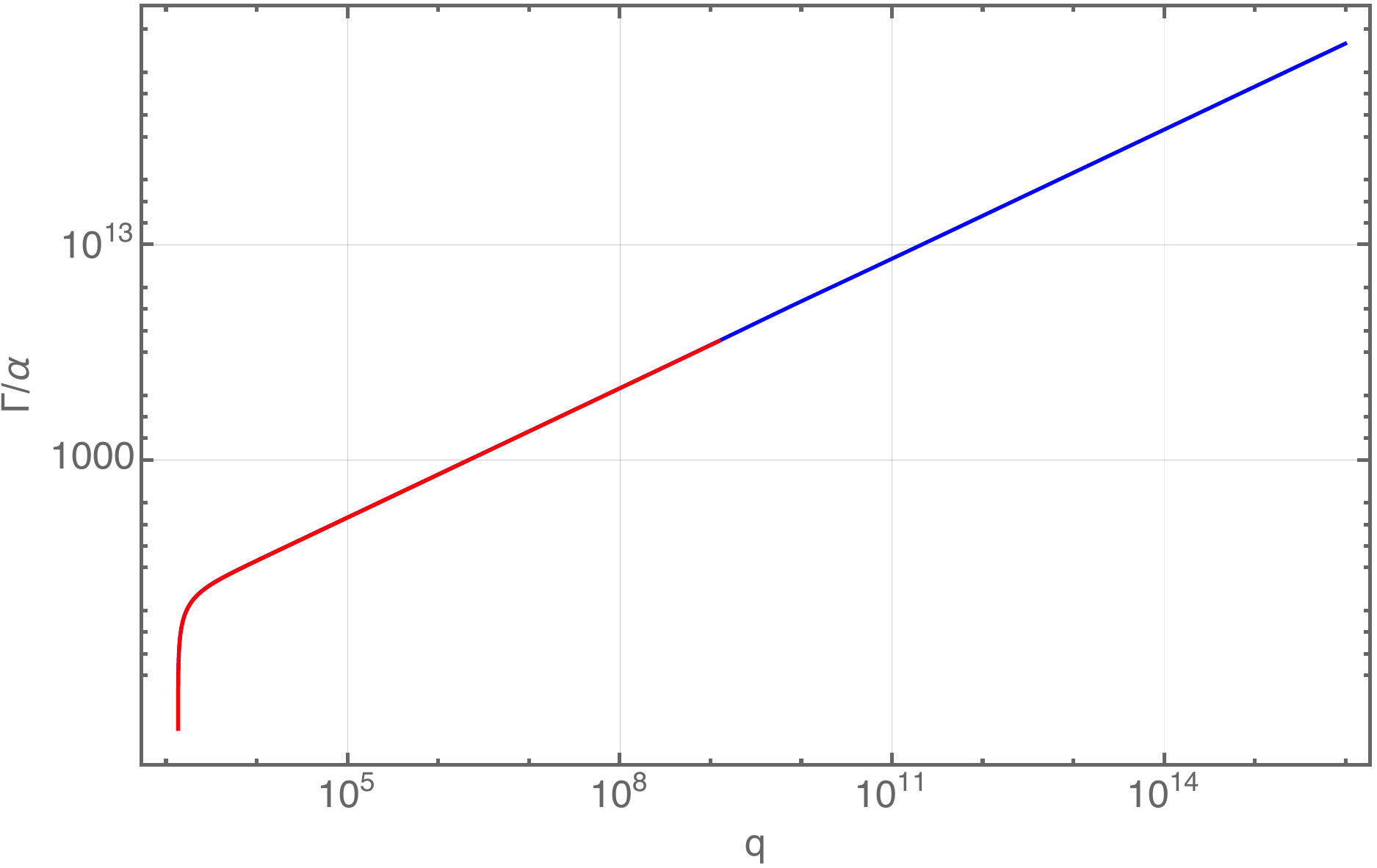}
		\caption{The same as Fig.~\ref{fig2} for the isotropic coefficients $\mathring{a}_0=\unit[10^{-10}]{GeV^{-1}}$ and $\mathring{a}_2=\unit[10^{-10}]{GeV^{-1}}$, respectively. One curve lies on top of the other.}
		\label{fig3}
	\end{figure}
	
	Finally, we proceed to study the effects caused by the isotropic coefficient $\mathring{a}_2$ of Eq.~\eqref{eq2.1b}. Since the latter is not contracted with time derivatives, there are no spurious dispersion relations. The positive-energy perturbative one reads
	\begin{equation}
		E_{\mathring{a}_2}=\sqrt{m_{\psi}^2+p^2}+\mathring{a}_2 p^2+\dots\,.
		\label{eq2.7}
	\end{equation}
	Comparing Eqs.~(\ref{eq2.3}) and (\ref{eq2.7}), we see that the former differs from the latter only by an additional fermion mass term. The threshold momentum has a behavior analogous to that of Eq.~\eqref{eq:threshold-a0}:
	\begin{equation}
		q_{\mathrm{th}}=\sqrt[3]{\frac{m^2_{\psi}}{4\mathring{a}_2}}+\dots\,.
	\end{equation}
	Let us look at the decay rates for $\mathring{a}_0$ and $\mathring{a}_2$, which are presented in Fig.~\ref{fig3}. The curves are indistinguishable. In contrast to the decay rate for $\mathring{m}_0$ in Fig.~\ref{fig2}, the curve for $\mathring{a}_2$ does not exhibit a knee. Note that the curve for $\mathring{a}_0$ again stops at a certain maximum momentum, where the exact expression for the energy becomes complex.
	
	\section{Constraints on the isotropic coefficients}
	\label{Sec5}
	
	If we assume that vacuum Cherenkov radiation is an expected phenomenon, it is possible to use data collected
	from UHECRs to put stringent bounds on the previous nonminimal, isotropic LV coefficients for the quark sector. To do so, we employ an effective description based on the modified QED introduced in Eq.~\eqref{eq:modified-QED} and put the electroweak and strong interactions aside.
	
	A primary cosmic ray observed on Earth has energy $E<E_{\mathrm{th}}$, where $E_{\mathrm{th}}$ is the fermion threshold energy following from the threshold momentum $q_{\mathrm{th}}$ of the previous sections. The latter can be expressed in generic form as $q_{\mathrm{th}}=\rho m_{\psi}^{\lambda}/\mathring{X}^{\sigma}$, where $\rho$, ${\lambda}$ and $\sigma$ are dimensionless parameters and $\mathring{X}$ is an isotropic coefficient. The inequality $E<E_{\mathrm{th}}$ leads us to
	\begin{equation}
		\mathring{X}<\left(\frac{\rho m_{\psi}^{\lambda}}{E}\right)^{\frac{1}{\sigma}}\,.
	\end{equation}
	Twice the experimental energy uncertainty $\Delta E$ is added to arrive at bounds with $2\sigma$ confidence level (CL):
	\begin{equation}
		\mathring{X}<\left(\frac{\rho m_{\psi}^{\lambda}}{E}\right)^{\frac{1}{\sigma}}+2\Delta E\left|\frac{\partial}{\partial E}\left(\frac{\rho m_{\psi}^{\lambda}}{E}\right)^{\frac{1}{\sigma}}\right|\,.
	\end{equation}
	We consider measurements reported by the Pierre-Auger observatory. In particular, the event
	737165 from Tab.~I in Ref.~\refcite{CosmicRays} is consulted, whose energy measured is $E=\unit[212\times 10^{18}]{eV}$ with an uncertainty of 25\%. The composition of the cosmic-ray primary is unknown, but we repeat the analysis done for the minimal-SME
	coefficients,\cite{Marco} where an iron nucleus is assumed to have been the primary. The energy of a single nucleon is $E/N$ with
	$N=56$ being the number of nucleons in the iron nucleus.
	
	\begin{table}[t]
		\centering
		\tbl{Constraints on isotropic coefficients in the u- and d-quark sectors at $2\sigma$ CL based on the event $737165$ of Ref.~\protect\refcite{CosmicRays}. \hfill\hfill}
		{\begin{tabular}{ccll}
				\toprule
				Fermion sector& Isotropic coefficient & Upper Constraint   \\
				\midrule
				u quark& $\mathring{m}_0$ & $< \unit[3\times 10^{-18}]{GeV^{-1}}$ \\
				& $\mathring{m}_2$  & $< \unit[3\times 10^{-18}]{GeV^{-1}}$ \\
				& $\mathring{a}_0$ & $<\unit[2\times 10^{-29}]{GeV^{-1}}$   \\
				& $\mathring{a}_2$ & $<\unit[2\times 10^{-29}]{GeV^{-1}}$\\
				\midrule
				d quark& $\mathring{m}_0$ & $< \unit[6\times 10^{-18}]{GeV^{-1}}$ \\
				& $\mathring{m}_2$  & $< \unit[6\times 10^{-18}]{GeV^{-1}}$  \\
				& $\mathring{a}_0$ & $<\unit[9\times 10^{-29}]{GeV^{-1}}$  \\
				& $\mathring{a}_2$ & $<\unit[9\times 10^{-29}]{GeV^{-1}}$ \\
				\bottomrule
		\end{tabular}}
		\label{tab1}
	\end{table}
	In addition, we assume that a Cherenkov photon is emitted by an up- or a down-quark of a nucleon. This quark carries a fraction $r=0.1$ of the nucleon energy and the quark masses considered are $m_{\mathrm{u}}\approx \unit[2.16\times 10^{-3}]{GeV}$ and $m_{\mathrm{d}}\approx \unit[4.70\times 10^{-3}]{GeV}$. The results of the constraints on the isotropic coefficients are presented in Tab.~\ref{tab1}. These bounds are better by many orders of magnitude than the ones obtained for some of the minimal coefficients in quarks.\cite{Marco} This is to be expected, since the effects of nonminimal coefficients increase with energy as compared to those of the minimal coefficients.

\end{document}